\documentclass[letterpaper,twocolumn,10pt]{article}

\usepackage{usenix}

\usepackage{filecontents}

\usepackage{graphicx}

\usepackage{enumitem}
\usepackage{color, soul}
\usepackage{xspace}
\usepackage{booktabs}
\usepackage{url}

\setlist[itemize]{nosep}

\usepackage{multirow}
\usepackage{multicol}

\usepackage{makecell}

\usepackage{amsmath}
\usepackage{amssymb}
\usepackage{bm}

\usepackage{tikz}
\newcommand*\emptycirc[1][1ex]{\tikz\draw (0,0) circle (#1);}
\newcommand*\halfcirc[1][1ex]{
  \begin{tikzpicture}
  \draw[fill] (0,0)-- (90:#1) arc (90:270:#1) -- cycle ;
  \draw (0,0) circle (#1);
  \end{tikzpicture}}
\newcommand*\fullcirc[1][1ex]{\tikz\fill (0,0) circle (#1);}

\newcommand{\threedots}{%
  \protect\begin{tikzpicture}[scale=0.1]
    \protect\filldraw (0,0) circle (15pt);
    \protect\filldraw (1.73,0) circle (15pt);
    \protect\filldraw (0.87,1.5) circle (15pt);
  \protect\end{tikzpicture}%
}

\begin{document}

\date{}

\title{\Large \bf The Code the World Depends On:\\A First Look at Technology Makers' Open Source Software Dependencies}

\author{
{\rm Cadence Patrick}\\
Spelman College
\and
{\rm Kimberly Ruth}\\
Stanford University
\and
{\rm Zakir Durumeric}\\
Stanford University
} 

\maketitle

\begin{abstract}
Open-source software (OSS) supply chain security has become a topic of concern for organizations. Patching an OSS vulnerability can require updating other dependent software products in addition to the original package. However, the landscape of OSS dependencies is not well explored: we do not know what packages are most critical to patch, hindering efforts to improve OSS security where it is most needed. There is thus a need to understand OSS usage in major software and device makers' products. Our work takes a first step toward closing this knowledge gap. We investigate published OSS dependency information for 108 major software and device makers, cataloging how available and how detailed this information is and identifying the OSS packages that appear the most frequently in our data.
\end{abstract}


\section{Introduction}

Software supply chain security has been the subject of increased attention in recent years. Although greater reliance on OSS has brought many benefits to organizations as an alternative to re-implementing functionality, it has also brought a greater need to patch OSS vulnerabilities. Patching an OSS vulnerability may involve not only the affected package itself, but also any other software products or packages that build upon it---a task that can be easier said than done. In 2021, the Log4Shell vulnerability in the popular Log4J open-source library was a wakeup call for the security industry, causing defenders to reckon with opaque layers of dependencies across a broad set of software products in order to remediate~\cite{log4j-challenges}. Supply chain attacks have continued to plague the industry since: a 2023 survey found that 10\% of organizations had a security breach due to open source vulnerabilities in the preceding 12~months~\cite{sonatype}. As companies' reliance on open source software continues to grow~\cite{sonatype}, the problem of managing vulnerable dependencies is likely to continue as well.

Despite laudable community efforts such as OpenSSF,\footnote{https://openssf.org} vulnerabilities in open source software continue to outpace defenders' capacity to address them. To address this growing challenge, CISA's Cyber Safety Review Board has recommended greater investment in securing open source software---specifically, focusing efforts on the packages that critical services are most reliant on~\cite{csrb-log4j}.

However, there is a major gap in our knowledge of what these most critical packages are. At the most basic level, we do not know which products rely on which OSS packages, and so we do not understand what ripple effects a particular vulnerability would cause. It is not enough to look at simple metrics such as number of GitHub stars to determine criticailty; instead, we need to understand the downstream consequences that a vulnerability in an OSS package would create in the broader software industry. By securing the OSS dependencies of major software and device makers, we in turn reduce risk for the critical infrastructure---hospitals, transportation, the electric grid, and more---that these products underpin.

This work aims to take a first step toward closing this knowledge gap. Specifically, we ask the following research questions:
\begin{itemize}
    \item How available is detailed OSS dependency data by major software and device companies?
    \item Which OSS packages are most frequently relied on by these major technology makers?
\end{itemize}

\noindent To that end, we search for published OSS dependency and licensing information for 108 major technology and device companies, investigating whether and how this information was made available. Although some companies do publish information about the open source packages they rely on, only 22.2\% of companies published a list of OSS package names that they depend upon. Among those that do publish this data, we count the most commonly used packages, including OpenSSL, \texttt{zlib}, and \texttt{ncurses}. In the process, we identify challenges with data availability and usability that we posit must be addressed if significant progress in this research area is to be made.

\section{Methodology}

We first chose 108 companies to investigate, focusing on major producers of software and devices used in critical infrastructure. We selected the top 10 companies by revenue from the 10 software-related industry categories in the 2023 Fortune 500 (Table~\ref{tab:f500}), supplemented with our own domain knowledge to include other prominent companies that did not appear in the Fortune 500 (e.g., because they are headquartered outside the United States).

Next, we conducted Google searches to look for OSS dependency information released by each company.\footnote{We used search keywords related to open source software licenses and legal information, such as ``open source software'' or ``OSS'' combined with ``license'', ``legal'', ``EULA'', ``documentation'', or ``third-party software''. We did not have success with ``SBOM'' as a search term.} Our search was conducted between June 28 and August 14, 2023. We collected information about each company’s use of third party software in any data format we found. Some companies released dependency lists on a per-product basis while others aggregated across their entire product suite; we took the union of all information we found for each company and deduplicated on OSS package names. We note that this produces a strict underestimate of software makers' OSS dependence, as there may be additional dependencies for products where OSS data was unavailable.

\begin{table}
    \small
    \centering
    \begin{tabular}{l c}
    \toprule
      Aerospace and Defense \\
      Computer Software \\
      Computers, Office Equipment \\
      Electronics, Electrical Equipment \\
      Industrial Machinery and ICS  \\
      Information Technology Services \\
      Internet Services and Retailing  \\
      Network and Other Communications Equipment  \\
      Semiconductors and Other Electronic Components  \\
      Telecommunications \\
    \bottomrule
    \end{tabular}
    \caption{Fortune 500 industry categories included in our analysis. We selected the top 10 companies in each category.}
    \label{tab:f500}
\end{table}

For the companies where we could not find any published OSS usage information, we contacted each company to ask for information about their OSS use, either through an OSS-specific email address where one was provided or else through a general customer service email.
Out of 67 companies, 13 replied via email, and none returned any new information. Reasons given in responses varied from legal barriers preventing the sharing of that information, to the companies not documenting the information to begin with.

Our OSS dependency data varied widely in format: PDFs containing tables or lists, web pages with bullet points, \texttt{.txt} files, and databases containing a combination of the above. We extracted lists using PyMuPDF and Tabula where possible and manually where not, standardized them into CSV format, and aggregated into one file. In the absence of a standard software identifier format in our data (reflecting a lack of consensus in the industry~\cite{cisa-identifiers, openssf-identifiers}), we cleaned and tokenized the data to find and count common strings. For this analysis, we did not analyze OSS packages on a per-version level; aggregating at the level of software package names better aligns with our goal of aiding prioritization for proactive vulnerability analysis and hardening efforts.

\paragraph{Limitations.} Our analysis is necessarily manual and thus cannot scale to the entire software industry; our results are constrained to the 108 major companies we consider, though we hope that many lessons we draw may be more broadly applicable. We have less visibility and expertise around non-US companies, so our analysis is US-leaning. Some companies may choose to release information about only the open-source dependencies whose licenses require it via attribution clauses; thus, we may not always know all dependencies for a company or product. It is possible that some companies have published information that is not indexed by Google Search, but given the lack of additional information provided upon request, we judge this to be unlikely. Some companies may choose to include packages they recursively depend on in their list of dependencies, while many may list only their direct dependencies; as this is distinction is generally not made clear in the documents we find, we do not distinguish in our analysis. Finally, much of the dependency information we find does not have a publication date listed and may be out of date. We treat the data we find as the best approximation we have.

\section{Results}

\begin{figure}
    \centering
    \includegraphics[width=\columnwidth]{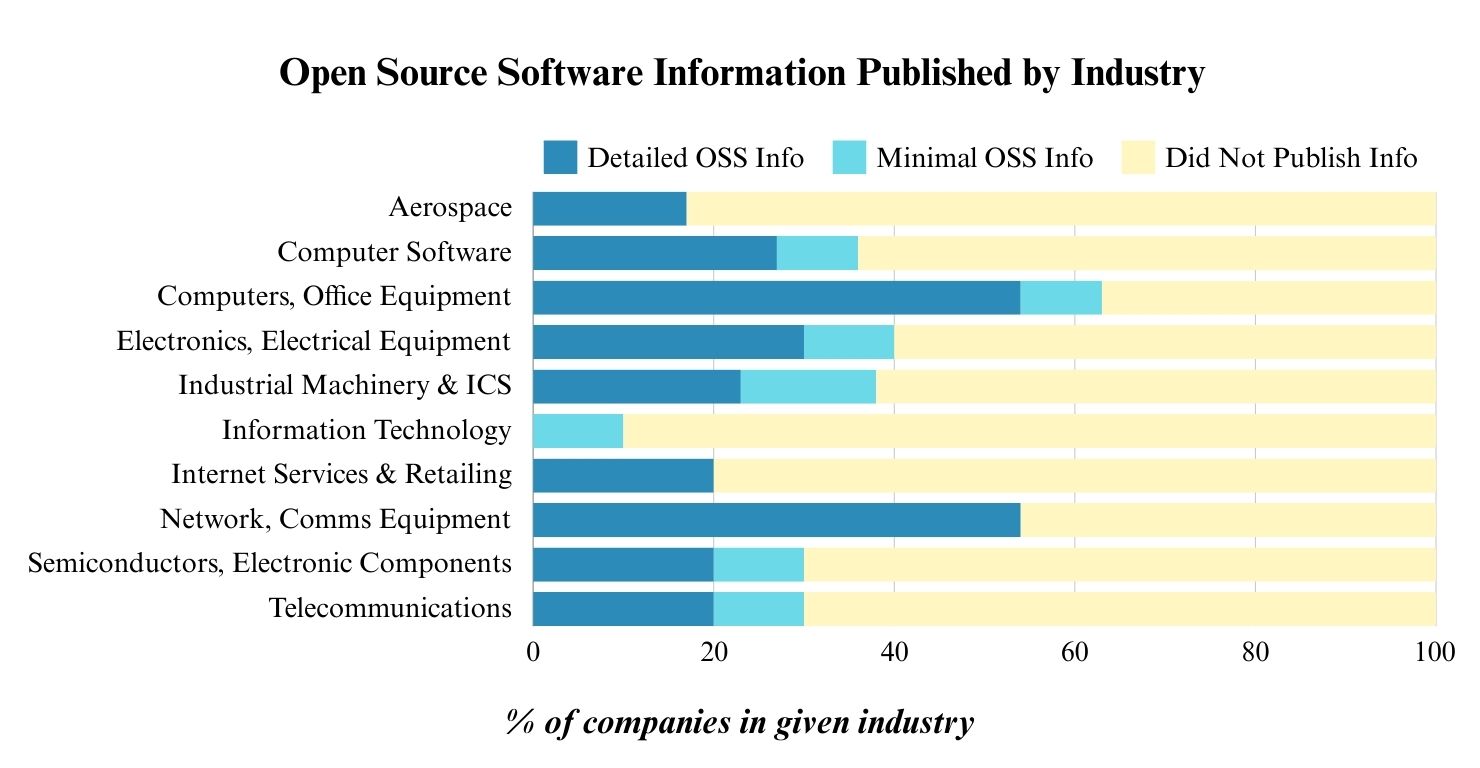} 
    \caption{Open source software dependency information availability by industry.}
    \label{fig:industrybarchart}
\end{figure}

Next, we describe the results of our analysis. We find that the companies we investigate have a mixed record of publishing OSS dependency data, with only 24 of 108~listing relied-upon OSS package names for any product, and we describe sector-level differences in data availability. We then analyze the published dependencies themselves, including how many OSS projects each company depends on and which packages are relied upon by the most of the software makers.

\subsection{OSS Dependency Data Availability}

\begin{table*}
    \small
    \centering
    \begin{tabular}{ccccc}\hline
    \toprule
      Industry & Companies & Total OSS Instances & Deduplicated & Industry Average \\
      \midrule
      \multirow{2}{*}{\makecell{Aerospace \\ and Defense}} & Airbus* & 90 & 90 & \multirow{2}{*}{53} \\ & Thales* & 36 & 17 &\\
      \hline
    \multirow{2}{*}{\makecell{Computer \\ Software}} & SAP* & 10 & 6 & \multirow{2}{*}{1181.5} \\
      & VMWare & 5451 & 2357 &\\
      \hline
    \multirow{5}{*}{\makecell{Computers, Office \\ Equipment}} & Dell & 882 & 322 & \multirow{5}{*}{181} \\ & Western Digital & 268 & 152 & \\ & Xerox Holdings & 441 & 378 & \\ & NetApp & 23 & 23 & \\ & Pitney Bowes & 21 & 21 & \\
    \hline
    \multirow{3}{*}{\makecell{Electronics, Electrical \\ Equipment}} & Honeywell & 42 & 42 & \multirow{3}{*}{1036.7} \\ & Zebra Technologies & 753 & 568 & \\
      & Vertiv Holdings & 4332 & 2500 &\\
      \hline
    \multirow{2}{*}{\makecell{Industrial Machinery \\ and ICS}} & Eaton* & 88 & 86 & \multirow{2}{*}{62} \\ & Schneider Electric* & 38 & 38 & \\
    \hline
    \multirow{2}{*}{\makecell{Internet Services \\ and Retailing}} & Google & 32 & 32 & \multirow{2}{*}{264.5} \\ & Meta Platforms & 526 & 497 & \\
    \hline
    \multirow{4}{*}{\makecell{Network and Other \\ Communications Equipment}} & Palo Alto Networks & 11262 & 1533 & \multirow{4}{*}{727.5} \\ & Juniper & 31 & 31 & \\ & Viasat & 197 & 184 & \\ & F5 & 1662 & 1162 &\\ 
    \hline
    \multirow{2}{*}{\makecell{Semiconductors and Other \\ Electronic Components}} & Intel & 25 & 25 & \multirow{2}{*}{25.5} \\
      & NVIDIA & 29 & 26 &\\
      \hline
    \multirow{2}{*}{Telecommunications} & Verizon & 6 & 6 & \multirow{2}{*}{18.5} \\ & Comcast & 32 & 31 & \\
    \bottomrule
    \end{tabular}
    \caption{OSS information by industry. We deduplicate OSS package counts across a company's multiple product lines as applicable. *Not in Fortune 500.}
    \label{tab:oss_per_org}
\end{table*}

We first examined how widely available OSS dependency data is. Although software makers have taken steps toward transparency in their OSS use, this data is still far from universally available. Our complete data availability results are detailed in Appendix~\ref{app:availability}.

Out of the 108 companies we investigated, 37 (34\%) had any open source software information published online across all 10 industries. The remaining 71 (66\%) appeared to have no information published that was relevant to our research. For the 37 companies publishing OSS dependency data, the available information varied widely in level of detail and in usefulness for our research questions. 24 companies (22.2\%) with published information list the names of software packages used for at least one product. The remainder only listed license text with no associated package names; although in some cases the package name can be inferred from the license (e.g., OpenSSL\footnote{Adobe's published document contains a license with the following text: ``The OpenSSL toolkit stays under a dual license, i.e. both the conditions of the OpenSSL License and the  original SSLeay license apply to the toolkit.''}), in many other cases it cannot (e.g., a generic GPL license text), and so we omit this data from further analysis as we do not want to bias our results based on what license the package is under.

As shown in Figure~\ref{fig:industrybarchart}, data availability also varies by industry sector.
The most forthcoming sector is Computers and Office Equipment, with 5 of 10 companies publishing usable OSS dependency data. The Network Equipment and Electronics sectors followed, with 4/10 and 3/10 in each sector publishing usable data, respectively.
The least forthcoming sector is IT Services, where no companies released usable OSS dependency information.

The product-level granularity of available data also differs between companies. Of the 37 companies publishing any information about OSS dependencies, 25 did so on a product-by-product basis, while the remaining 12 aggregated information across their entire product suite. While per-product breakdowns are helpful for defenders to know which specific components do or do not need to receive security updates, we note that not all companies release information for all of their products: 15 of 25 had information for all products linked from a single database, while the remaining 10 only had per-product information available ad hoc for a subset of their products.

Although analyzing packages at the granularity of version numbers is out of scope for the current investigation, we also note that of the 24 companies listing package names they depend on, only 11 include version numbers for all packages, and 9 include no version numbers at all. With 10.2\% (11/108) of companies releasing OSS dependency data at version-level granularity, future research into companies' dependence on vulnerable packages will need to contend with both data freshness and data availability limitations.

\subsection{Contents of OSS Dependency Lists}

Even with only a small fraction of companies publishing OSS information, the available data shows a rich ecosystem of OSS dependence. Table~\ref{tab:oss_per_org} shows the scale of OSS dependence for the 24 companies with usable data.
Across these 24 companies, we observe over 10K 
total instances of OSS use; even after deduplicating packages across each company's product lines where applicable, we see an average of 422 dependencies per company.
There is high variance in the number of observed dependencies ($\sigma=710.9$), ranging from SAP and Verizon with 6 unique packages reported to VMWare and Vertiv with well over 2,000~dependencies.
We underscore that this is a strict underestimate of OSS dependence, since for some companies we only have dependency information for a single product, and OSS dependence across a company's multiple product lines may be even greater. Furthermore, due in part to limited available data on software versions, we do not analyze dependencies recursively, meaning there are likely even more dependencies lying deeper in the dependency tree.

OSS dependence varies by industry sector. As shown in Table~\ref{tab:oss_per_org}, the highest volume of OSS usage occurs in the computer software, electronics equipment, and network equipment companies we examined, primarily driven by the heavy-hitters VMWare, Vertiv, Palo Alto Networks, and F5. 
Although we caution against overgeneralization due to high variance among the fairly small number of companies publishing OSS data, we hypothesize that at least some of the differences we see may arise due to factors such as differing internal policies about OSS use, differing capacities and priorities of in-house development teams, or differing sets of software needs which the OSS ecosystem can help to fill. We encourage future work to further explore such differences.
Turning to the set of package names depended upon, we observe a long tail of OSS dependencies: there are over 8000 unique package names in our dataset.
Table~\ref{tab:popular_oss} shows the 27 OSS packages depended upon by at least 7 companies. The most relied upon dependencies fall broadly into three groups:

\vspace{0.05in}\noindent\textbf{Network and cryptographic protocols}: Common network protocol dependencies include secure communication (\texttt{openssl} and \texttt{openssh}), local networking (\texttt{dhcp} and \texttt{net-snmp}), network traffic capture and analysis (\texttt{libpcap} and \texttt{tcpdump}), and layer-7 data transfer libraries (\texttt{libcurl} and \texttt{okhttp}). Notably, we confirmed that \texttt{dhcp} refers to the ISC implementation of DHCP\footnote{\url{https://www.isc.org/dhcp/}} in every instance it appears in our dataset; ISC declared its DHCP implementation to be no longer maintained as of late 2022, and yet 7 companies' documentation still states that they rely upon it.

\vspace{0.05in}\noindent\textbf{File and system utilities}: These include \texttt{zlib} and \texttt{bzip2} for compression, \texttt{expat} and \texttt{libxml2} for XML parsing, \texttt{libpng} and \texttt{gson} for PNG and JSON handling, and several Linux kernel and system utilities such as \texttt{u-boot}, \texttt{e2fsprogs}, and \texttt{kmod}.

\vspace{0.05in}\noindent\textbf{Programming tools and frameworks}: We frequently see \texttt{jquery} and \texttt{boost} for streamlining web and C++ programming, respectively; \texttt{ncurses} and \texttt{libevent} for developing text-based UI and event-driven programs, respectively; and the standalone scripting language \texttt{bash}.

\vspace{0.05in}\noindent OpenSSL is the most relied upon package in our data, appearing in 15 companies' dependency lists, with zlib a close second at 13 companies. OpenSSH, ncurses, and libpcap follow with 9 appearances each.


        


\begin{table}
    \footnotesize
    \centering
    \begin{tabular}{cp{2in}r}\toprule
       OSS & Description & \# Orgs \\
       \midrule
        openssl & cryptographic protocol library & 15 \\
        zlib & data compression library & 13 \\
        ncurses & library for text-based UI & 9 \\
        openssh & encrypted communications library & 9 \\
        libpcap & network traffic capture library & 9 \\
        jquery & web programming library & 8 \\
        okhttp & performant HTTP library & 8 \\
        libevent & event-driven programming library & 8 \\
        libxml2 & XML parser library & 8 \\
        bzip2 & file compression utility & 8 \\
        e2fsprogs & file system administration utilities & 8 \\
        expat & XML parser & 8 \\
        kmod & Linux kernel module manager & 8 \\
        libpng & PNG utility library & 8 \\
        tcpdump & network packet analyzer & 8 \\
        u-boot & bootloader for embedded Linux systems & 7 \\
        util-linux & multi-purpose Linux package & 7 \\
        net-snmp & network management protocol library & 7 \\
        libcurl & network data transfer library & 7 \\
        ethtool & network interface device manager & 7 \\
        boost & multi-purpose C++ library & 7 \\
        busybox & embedded Unix software suite & 7 \\
        bash & scripting language & 7 \\
        freetype & font and text renderer & 7 \\
        pcre & expressive regex library & 7 \\
        gson & Java JSON serialization library & 7 \\
        dhcp & ISC network management protocol library & 7 \\
        \bottomrule
    \end{tabular}
    \caption{Most Popular OSS Dependencies}
    \label{tab:popular_oss}
\end{table}

\section{Discussion and Conclusion}

Attacks on OSS dependencies are a significant concern for software and hardware makers. Contributors in the open source community have already been making admirable efforts to secure the open source ecosystem; however, the sheer scale of open source code today presents significant challenges in scaling up security to match. By identifying the projects that are most relied upon by major producers of software, we can better direct resources and effort toward critical components of our tech ecosystem.

A crucial precursor to that goal, however, is assembling data from software makers on their use of OSS. Today, there is no universal expectation for companies to publish such information---and indeed, some do not track this information at all. We commend those companies that choose to release this data publicly. Although we are optimistic that some are beginning to do so, we need a continued push for transparency. In addition, standardizing the format of released data would go a long way toward facilitating analysis across the entire industry. Recent efforts to increase Software Bill of Materials (SBOM) adoption~\cite{cisa-sbom, osti-sbom} are positive steps forward, although the lack of publicly released SBOMs hinders ecosystem-wide analysis, and many other challenges remain~\cite{xia2023sboms, stalnaker2024sboms}.

Our findings are only a first step in the analysis of critical software makers' dependencies. We examined OSS projects that companies directly reported that they depend on, but we did not trace these dependencies recursively. In part this is because OSS version numbers were not always provided in the data, limiting our ability to reliably trace dependencies further. Future work, however, should aim to identify highly relied upon packages lying deeper in the dependency tree. Beyond that, other future work may aim to better time-stamp dependency data and conduct deeper investigations into companies' reliance on vulnerable or end-of-life software packages.

We envision a future in which security events like Log4Shell no longer occur: one in which widely depended upon packages receive enhanced security hardening, and in which the broader impact of a vulnerability is well understood and anticipated to make remediation easier in the rare case of an event. We hope that this work serves as a catalyst for future data sharing, synthesis, and ultimately security analysis. There is still much more to be done to shed light on OSS dependencies.

\section*{Acknowledgments}

We thank Breauna Spencer and the other Stanford LINXS program coordinators and participants for their support and helpful conversations. This work is funded in part by the Stanford LINXS program and by an NSF Graduate Research Fellowship DGE-1656518.

\bibliographystyle{plain}
\bibliography{refs}

\appendix
\section{Data Availability Details}
\label{app:availability}

In Table~\ref{tab:circle_table}, we detail OSS dependency data availability across all the companies we investigated who had at least some information available (37 out of 108).

\begin{table*}
    \small
    \centering
    \begin{tabular}{c c c c c c}
        \toprule
        Industry & Company & Product Coverage\textsuperscript{a} & \makecell{Product-Level \\ Granularity\textsuperscript{b}} & \makecell{Package Names \\ Present\textsuperscript{c}} & \makecell{Versions \\ Present\textsuperscript{d}} \\
        \midrule
        \multirow{2}{*}{\makecell{Aerospace \\ and Defense}} & Airbus & \fullcirc & \emptycirc & \fullcirc & \fullcirc \\
            & Thales & \fullcirc & \threedots & \fullcirc & \emptycirc \\
            \hline \noalign{\vskip 1mm}
        \multirow{4}{*}{\makecell{Computer \\ Software}} & Oracle & \halfcirc & \emptycirc & \emptycirc & \emptycirc \\ 
        & Adobe & \fullcirc & \emptycirc & \emptycirc & \emptycirc \\ 
        & SAP & \halfcirc & \threedots & \fullcirc & \fullcirc \\
            & VMWare & \halfcirc & \threedots & \fullcirc & \fullcirc \\
            \hline \noalign{\vskip 1mm}
        \multirow{4}{*}{\makecell{Computers, Office \\ Equipment}} & Dell & \halfcirc & \threedots & \fullcirc & \emptycirc \\
        & HP & \fullcirc & \emptycirc & \emptycirc & \emptycirc \\ 
        & Western Digital & \fullcirc & \threedots & \fullcirc & \halfcirc \\ 
            & Xerox Holdings & \fullcirc & \threedots & \fullcirc & \emptycirc \\
            & NetApp & \halfcirc & \threedots & \fullcirc & \emptycirc \\ 
            & Pitney Bowes & \halfcirc & \threedots & \fullcirc & \emptycirc\\
            \hline \noalign{\vskip 1mm}
        \multirow{3}{*}{\makecell{Electronics, Electrical \\ Equipment}} & Honeywell & \halfcirc & \threedots & \fullcirc & \fullcirc \\ 
        & Rockwell Automation & \fullcirc & \emptycirc & \emptycirc & \emptycirc \\ 
        & Zebra Technologies & \fullcirc & \threedots & \fullcirc & \halfcirc \\
              & Vertiv Holdings & \fullcirc & \threedots & \fullcirc & \halfcirc \\
              \hline \noalign{\vskip 1mm}
        \multirow{4}{*}{\makecell{Industrial Machinery and ICS}} & Otis & \fullcirc & \emptycirc & \emptycirc & \emptycirc \\
            & Wabtec & \fullcirc & \emptycirc & \emptycirc & \emptycirc \\
            & Eaton & \fullcirc & \emptycirc & \fullcirc & \fullcirc \\
            & Schneider Electric & \halfcirc & \threedots & \fullcirc & \fullcirc \\
            & Siemens & \fullcirc &\threedots & \emptycirc & \emptycirc \\
            \hline \noalign{\vskip 1mm}
        \makecell{IT Services} & IBM & \fullcirc & \emptycirc & \emptycirc & \emptycirc \\
            \hline \noalign{\vskip 1mm}
        \multirow{2}{*}{\makecell{Internet Services \\ and Retailing}} & Google & \halfcirc & \threedots & \fullcirc & \emptycirc \\
        & Meta Platforms & \fullcirc & \threedots & \fullcirc & \emptycirc\\
            \hline \noalign{\vskip 1mm}
        \multirow{7}{*}{\makecell{Network and Other \\ Communications Equipment}} & Cisco & \fullcirc & \threedots & \emptycirc & \emptycirc \\ 
        & Arista & \fullcirc & \threedots & \emptycirc & \emptycirc \\ 
        & Palo Alto Networks & \fullcirc & \threedots & \fullcirc & \fullcirc \\ 
            & Juniper & \fullcirc & \threedots & \fullcirc & \fullcirc \\
            & Viasat & \fullcirc & \threedots & \fullcirc & \emptycirc \\
            & F5 & \fullcirc & \emptycirc & \fullcirc & \fullcirc \\ 
            & Netgear & \fullcirc & \threedots & \emptycirc & \emptycirc \\
            \hline \noalign{\vskip 1mm}
        \multirow{3}{*}{\makecell{Semiconductors and Other \\ Electronic Components}} & Intel & \fullcirc & \emptycirc & \fullcirc & \halfcirc \\
            & Broadcom & \fullcirc & \emptycirc & \emptycirc & \emptycirc \\
            & NVIDIA & \halfcirc & \threedots & \fullcirc & \fullcirc \\
            \hline \noalign{\vskip 1mm}
        \multirow{3}{*}{\makecell{Telecommunications}} & Verizon & \halfcirc & \threedots & \fullcirc & \fullcirc \\
        & Comcast & \fullcirc & \threedots & \fullcirc & \emptycirc \\
        & AT\&T & \fullcirc & \threedots & \emptycirc & \emptycirc \\
        \bottomrule
    \end{tabular}
    \caption{Data availability for companies with at least some dependency information public. The other 71 companies had no data available at all. \\
    \textsuperscript{a} \fullcirc \,= dependencies available for all products; \halfcirc \,= dependencies available for only some products \\
    \textsuperscript{b} \protect{\threedots} = dependencies available at a per-product granularity; \emptycirc \,= all products' dependencies aggregated into one list \\
    \textsuperscript{c} \fullcirc \,= names of packages available; \emptycirc \,= only license text available \\
    \textsuperscript{d} \fullcirc \,= version numbers available; \halfcirc \,= some version numbers available; \emptycirc \,= no version numbers available
    }
    \label{tab:circle_table}
\end{table*}

\end{document}